\documentclass[12pt]{article}
\usepackage{amsmath,amssymb,amsthm,amsxtra,overpic,bbm,bm,epsfig,ulem}
\usepackage{color}
\usepackage{cite}

\usepackage{hyperref}
\usepackage{url}

\usepackage{slashed}
\def \bea {\begin{eqnarray}}
\def \eea {\end{eqnarray}}

\def\thefootnote{\fnsymbol{footnote}}

\begin{document}

\vspace{0.2cm}

\begin{center}
{\large\bf  Quantum field theory approach to  neutrino oscillations in dark matter and  implications at JUNO}
\end{center}

\vspace{0.2cm}

\begin{center}
{\bf Wei Chao }$^{1,}$
\footnote{E-mail: chaowei@bnu.edu.cn}
\\
{\small $^{1}$Key Laboratory of Multi-scale Spin Physics, Ministry of Education, Beijing Normal University, Beijing 100875, China} \\
{\small $^{2}$Center of Advanced Quantum Studies, School of Physics and Astronomy, Beijing Normal University, Beijing, 100875, China} 
\end{center}

\vspace{1cm}

\begin{abstract}
Neutrino oscillation is a significant physical process worthy of in-depth exploration. In this paper, we investigate the matter effect of massive neutrinos in a scalar-type ultra-light dark matter and calculate the neutrino oscillation probability using the quantum field theory method. The result reveals that the neutrino oscillation probability derived from the quantum field theory approach exhibits no additional time dependence, which marks the most significant distinction from the oscillation result obtained through the quantum mechanics method. Furthermore, we discuss predictions of the Juno experiment regarding neutrino oscillation behavior in scalar-type ultra-light dark matter. This study extends the understanding of the interaction between neutrinos and dark matter, which warrants further exploration.

\end{abstract}

\newpage

\def\thefootnote{\arabic{footnote}}
\setcounter{footnote}{0}

\section{Introduction}

Atmospheric, solar, reactor and accelerator neutrino oscillations have confirmed that neutrinos are massive and lepton flavors are mixed with each other. Strengths of lepton flavor mixing and CP violation in the charged-current (CC) interactions of the standard model (SM)  can be described by the $3\times 3$  Pontecorvo-Maki-Nakagawa-Sakata (PMNS) matrix~\cite{Pontecorvo:1957cp,Maki:1962mu} ${\cal U}$, which can be  parameterized as 
\begin{equation}
{\cal{U}} =
\begin{pmatrix}
c_{12} c_{13} &
s_{12} c_{13} &
s_{13} e^{-i\delta_{CP}} \\[6pt]
- s_{12} c_{23} - c_{12} s_{23} s_{13} e^{i\delta_{CP}} &
c_{12} c_{23} - s_{12} s_{23} s_{13} e^{i\delta_{CP}} &
s_{23} c_{13} \\[6pt]
s_{12} s_{23} - c_{12} c_{23} s_{13} e^{i\delta_{CP}} &
- c_{12} s_{23} - s_{12} c_{23} s_{13} e^{i\delta_{CP}} &
c_{23} c_{13}
\end{pmatrix}
\,
\label{eq:UPMNS}
\end{equation}
where $s_{ij} = \sin\theta_{ij}$, $c_{ij} = \cos\theta_{ij}$ and $\delta_{CP}$ is the Dirac CP-violating phase. Mixing angles $\theta_{12}^{}$, $\theta_{13}$ and $\theta_{23}$  are mainly measured by solar neutrino experiments, reactor neutrino experiments, atmospheric and long-baseline neutrino experiments, respectively. Next generation neutrino oscillation experiments aim to determine the neutrino mass ordering and the Dirac CP phase~\cite{DeSalas:2018rby}. 

It should be mentioned that neutrino oscillation experiments can also be taken as a facility for detecting new physics beyond the SM, which could be non-standard neutrino interactions~\cite{Ohlsson:2012kf}, sterile neutrinos~\cite{Shi:1998km,Abazajian:2012ys}, unitarity violation in the PMNS matrix~\cite{Antusch:2006vwa}, Lorentz Invariance Violation~\cite{Colladay:1998fq}, et cetera. Matter effects~\cite{Wolfenstein:1977ue,Wolfenstein:1979ni,Mikheyev:1985zog,Mikheev:1986wj}, which are usually described by the Wolfenstein potential and are caused by coherent forward scattering off various matters, are particularly important in detecting parameters of neutrino physics and also new physics. For example, DUNE~\cite{DUNE:2020ypp} will make use of matter effects to determine the neutrino mass hierarchy. The Wolfenstein potential induced by CC and neutral-current (NC) interactions have been dedicatedly studied, yet matter effects arising from the interactions between neutrinos and exotic new physics remain under-explored. Considering that we are in the era of precision neutrino physics and oscillation parameters can be measured at the sub-percent level, matter effects induced by new physics should not be overlooked.

In this paper, we study the matter effect induced by interactions between neutrinos and ultralight dark matter (DM)~\cite{Bertone:2004pz}. Astronomical and cosmological observations have confirmed the existence of DM, which accounts about 85\% of the total mass of the universe, and galaxies are embedded in DM halos. The ambiguity in the particle properties of DM has given rise to numerous DM candidates~\cite{Cirelli:2024ssz}, whose masses span a range of over 90 orders of magnitude. Among these, fuzzy DM~\cite{Hu:2000ke,Li:2013nal,Hui:2016ltb,Arias:2012az}, studied in the context of dark photon~\cite{Fabbrichesi:2020wbt}, axion~\cite{Preskill:1982cy} and scalar fields~\cite{Irsic:2017yje,Chao:2024owf}, is regarded as attractive wave-like DM candidate. These ultralight DM can resolve puzzles related to structure formation in the Universe~\cite{Hu:2000ke,Klypin:1999uc,Boylan-Kolchin:2011qkt} and can be detected in  laboratory experiments, such as cavity-based haloscopes~\cite{Sikivie:1983ip} or helioscopes~\cite{Irastorza:2011gs}, LC circuits,~\cite{Sikivie:2013laa}  ground based interferometer~\cite{Abramovici:1992ah}, etc. If neutrinos couple to the  ultralight DM, there will be a new Wolfenstein potential in neutrino oscillations. Measuring this new matter effect will provides indirect evidences for the existence of  DM. Historically, this effect has been described  by the quantum mechanical approach~\cite{Brdar:2017kbt,Krnjaic:2017zlz,Liao:2018byh,Chao:2020bti,Ge:2018uhz}.   In this paper, we revisit  this effect using the quantum field theory approach~\cite{Cardall:1999ze,Beuthe:2001rc,Dolgov:2002wy,Giunti:2002xg,Kobzarev:1980nk,Giunti:1993se,Grimus:1996av,Grimus:1998uh,Beuthe:2002ej,Dolgov:2005nb,Kopp:2009fa,Akhmedov:2010ms,Dvornikov:2024zow,Dvornikov:2025mdu,Morozumi:2025gmw,Grimus:2019hlq,Dvornikov:2025dam}, which has been proved to be a useful tool in understanding the nature of neutrino oscillation.   We focus on the scalar-type ultralight DM case in this paper and calculate the propagator of neutrinos in the background  of DM.  Then we derive the two-flavor neutrino oscillation probability and compare our results with that derived from the quantum mechanic approach.  Finally, we study this new matter effect in the JUNO neutrino oscillation experiment. 

The remaining of the paper is organized as follows: In section II we give a brief introduction to the scalar type ultralight DM and derive the neutrino propagator in the DM background. Section III is focused on the calculation of neutrino oscillation probability and  phenomenological simulations. The last part is concluding remarks.

\section{Neutrino propagators in DM}
In this section we calculate propagators of massive neutrinos in the external field.  We will not consider traditional matter effect, which has been discussed in Refs.\cite{Wolfenstein:1977ue,Wolfenstein:1979ni,Mikheyev:1985zog,Mikheev:1986wj,Huang:2023nqf,Huang:2024rfb} and references cited therein. Instead, we consider the propagator of Majorana neutrino in fuzzy DM~\cite{Gan:2025nlu}, which is taken as a scalar field $\varphi$.  The Lagrangian is given by 
\begin{eqnarray}
{\cal L} &=& \overline{\nu^{\alpha}_{L }} i \gamma^\mu \partial_\mu \nu_{L}^{ \alpha} - \frac{1}{2} {\cal G}_{\alpha \beta}^{} \varphi  \overline{\nu_L^\alpha }  \nu_L^{\beta} - \frac{1}{2} \overline{\nu_{L\alpha }^{  C }}  m^{\alpha \beta }_{0}    \nu_{L\beta}^{}  + {\rm h.c.}  \nonumber \\
&=&\frac{1}{2} \overline{\nu^\alpha} \left( i \delta_{\alpha \beta}^{}  \gamma^\mu \partial_\mu^{}  - m_{0, \alpha \beta}^{} -{\cal G}_{\alpha \beta}^{} \varphi \right) \nu^{\beta} 
\end{eqnarray}
where $m_0$ is the $3\times3$ symmetric neutrino mass matrix induced by the seesaw mechanism,  $g_{\alpha \beta}^{}$ is the effective coupling between active neutrinos and the DM and is taken as real symmetric matrix, $\nu$ is defined as $\nu=\nu_L + \nu_L^C$.  The ultralight scalar DM can be treated as classical field, and the solution to the classical equation of motion in the nonrelativistic limit can be written as~\cite{Berlin:2016woy}
\bea
\varphi (x) =\frac{\sqrt{\rho_\varphi (x)}}{m_\varphi} \cos \left( m_\varphi t - m_\varphi \vec v \cdot \vec x\right)
\eea
where $\rho_\varphi \sim 0.3 ~{\rm GeV/cm^3}$ being the local DM density, $m_\varphi$ is the DM mass, $\vec v$ is the DM velocity. Considering that DM is cold, we neglect the spatial variation of $\varphi$ in our solar system.  The Fourier transformation of the DM is 
\bea
\varphi (p) \approx 2\pi \sqrt{\rho_\varphi } \delta^{} (p_0^2 -m_\varphi^2) \delta^{(3)} (\vec p)
\eea
which will be applied to the calculation of the neutrino propagator. The mass matrix $m_{0}$ can be diagonalized by the unitary transformation, ${\cal U}^T m_0^{} {\cal U}^{} ={\rm Diag} \{ m_1, m_2, m_3 \}$, where ${\cal U}$ is the unitary matrix given in Eq.~(\ref{eq:UPMNS}), After the diagonalization, the equation of motion for the mass eigenstates take the following form
\begin{eqnarray} 
\left(i \gamma_\mu \partial^\mu   - m_i^{}  \right) \hat \nu_i   =  \sum_{j  } \bar {\cal G}_{ij}^{} \varphi^{}  \hat \nu_j^{} 
\end{eqnarray}
where $\bar {\cal G} = {\cal U}^T {\cal G} {\cal U}^{}$ and $\hat \nu = {\cal U}^\dagger \nu $.  The Feynman propagator for free left-handed active neutrinos can be written as~\cite{Fukugita:2003en} 
\begin{eqnarray}
S_F^i(x) =\int \frac{d^4 p}{(2\pi)^4} e^{-i p\cdot x} \frac{i}{2} \left( 1- \frac{\vec \sigma \cdot \vec p}{E_i} \right) \frac{1}{p_0 -E_i + i\varepsilon} \label{master0}
\end{eqnarray}
where $E_i =\sqrt{p^2 + m^2_{\rm i }}$. When only considering the effect of $\bar {\cal G}_{ii} \varphi$,  the Feynman propagator for $\hat \nu_i$ is the same as that in the Eq. (\ref{master0}) up to the replacement, $m_i \to \tilde m_i =m_i + \bar {\cal G}_{ii} 2\pi \sqrt{\rho }/m_\varphi$. We follow the strategy presented  in the section 2-5-2 of the Ref~\cite{Itzykson:1980rh} to derive the full propagator in the existence of nonzero $\bar {\cal G}_{ij}^{}$ with $i \neq j$,
\bea
S_A^i(x_f, x_i) &=& S_F^i (x_f, x_i)+ \iint  d^4 x_1 d^4 x_2 S_F^i (x_f -x_1) g_{ij} \varphi(x_1)S_F^j (x_1, x_2 ) g_{ji} \varphi (x_2) S_F^i (x_2, x_i)
\nonumber \\ && + \cdots
\eea
For Simplicity, we assume that $\bar {\cal G}_{12} =\bar {\cal G}_{21} \neq 0$ and  all other off-diagonal terms are zero.
Then $S_A^{1\to 1} (p)$ can be written as
\bea
S_A^{1\to 1} (p ) & =& S_F^i (p) + \iint d^4 p_1 d^4 p_2 S^i_F( p+ p_1 + p_2) g_{ij}^{} \varphi(p_1) S_F^j (p_2 + p  )  g_{ji}^{} \varphi(p_2)  S^i_F(p) + \cdots \nonumber \\
&\approx& S_F^i (p) \sum_{k=0}^{}\left[    (2 \pi \bar {\cal G}_{12}  \sqrt{\rho_\phi} /m_\varphi^{} ) S_F^j (p)  (2 \pi \bar {\cal G}_{12}  \sqrt{\rho_\phi} /m_\varphi^{} )  S_F^i (p)  \right ]^k \nonumber \\
&\approx& \frac{ i p\cdot \sigma }{2p_0 }\frac{p_0 -E_2}{ (p_0 -E_1+ i\varepsilon)(p_0-E_2+ i\varepsilon) - \kappa^2  } \equiv \frac{ i p\cdot \sigma }{2p_0 } \Sigma^{11}
\eea
where the approximation in the second line arises from the consideration of the smallness of $m_\varphi$, such that $S_F^i (p + p_k ) \sim S_F^{i} (p)$ when performing integration to the momenturm $p_k$, $E_i =\sqrt{\vec p^2 + \tilde m_i^2}$ and $\kappa=2 \pi \bar {\cal G}_{12}  \sqrt{\rho_\phi} /m_\varphi^{} $. 
The propagator for the second neutrino mass eigenstate takes the similar form
\bea
S_A^{2\to 2} (p) =\frac {(p_0 -E_1)}{ (p_0 -E_1+ i\varepsilon)(p_0-E_2+ i\varepsilon) - \kappa^2  } \equiv \frac{ i p\cdot \sigma }{2p_0 } \Sigma^{22}
\eea
The propagator between the first two neutrino mass eigenstate can be written as 
\bea
S_A^{1\to 2}(p) = \frac{ i p\cdot \sigma } {2 p_0} \frac{\kappa }{  (p_0 -E_1+ i\varepsilon )(p_0-E_2+ i\varepsilon ) -\kappa^2 } \equiv  \frac{ i p\cdot \sigma } {2 p_0} \Sigma^{12}
\eea
and $S_A^{2\to 1}$ takes the same form. One can check that these propagators will be turned into the standard form when turning off DM-neutrino interactions.  Using these propagators, we can calculate the neutrino oscillation probability, which will be done in the next section.

\section{Neutrino oscillation probability}
In quantum field theory, the neutrino oscillation is described by the $2\to 2$ Feynman diagram, by considering neutrino production, propagator and detection as a single process.  The states that describe the particles accompanying neutrino production and detection can be written as~\cite{Beuthe:2001rc}
\bea
|\psi_{i } (x)\rangle = \int \frac{d^3 p }{(2\pi)^3 \sqrt{2 E_i(p)}} e^{-i p\cdot x} f_{\Psi_i} (p,P) |p \rangle 
\eea 
where $|p\rangle$ is one-particle momentum eigenstate with momentum $p$, $E_i =\sqrt{m_i^2 + p^2}$, $f(p, P)$ is the distribution function with $P$ the mean momentum.  The amplitude for the neutrino oscillation can be written as
\bea
i T_{\alpha \to \beta} = \left\langle P_f, ~D_f \left|  \frac{(-i)^2}{2 !}T \left[ \int d^4 x d^4 y {\cal L}_{CC} (x) {\cal L}_{CC}(y) \right] \right |P_i , ~D_i \right \rangle  
\eea
where ${\cal L}_{CC}$ is the charged current interaction, $P_i $, $P_f$ are initial and final states during production, $D_i$ and $D_f$ are states accompanying neutrino detections. A direct calculation results in 
\bea
i T_{\alpha \to \beta} = \sum_{ij} {\cal U}^*_{\alpha i} {\cal U}_{\beta j}^{}  \int {d^3 q \over (2\pi)^3} {d^3 k\over (2\pi)^3}{ d^3 q^\prime\over (2\pi)^3}{ d^3 k^\prime  \over (2\pi)^3} f_{P_i} (q, Q) f_{P_f}^* (k, K) f_{D_i} (q^\prime , Q^\prime ) f_{D_f} (k^\prime , K^\prime ) \nonumber \\
\times\int d^4 x d^4 y e^{- i(q^\prime -k^\prime)\cdot (y+x_P) }  \hat M_D^{}( q^\prime, k^\prime) \int \frac{d^4 p }{(2\pi)^4} S_A^{i \to j}(p) e^{-i p\cdot (y-x)}\hat M_P^{}(q, k) e^{- i (q-k) \cdot (x+x_D) } \label{master12}
\eea
where $\hat M_{P,D}$ are plane wave amplitudes  for neutrinos in production and detection processes with neutrino spinors excluded. Neutrino spinors can be extracted from the neutrino propagator, i.e.,  $p\cdot \sigma/(2p_0)$ can be  replaced with spinor product, $ u(p,-) u^\dagger (p, -)/(2p_0)$, and can be absorbed by $\hat M_{P,D}$ to form true amplitudes $M_{P,D}^{}$, with $M_P= \bar u (p, -) \hat M_P/\sqrt{2p_0}$ and $M_D = \hat M_D u(p, -)/(2p_0)$. Performing integrations to $x $ and $y$,   terms in the second line of the  Eq. (\ref{master12}) can be written as
\bea
\int \frac{d^4 p} {(2\pi)^4} \Sigma^{i \to j} (p ) e^{-i p\cdot (x_D-x_P)} (2\pi)^4 \delta^4 ( q^\prime -k^\prime + p) (2\pi)^4 \delta^4 ( q -k- p)  {\cal M }_{i P}^{} {\cal M}_{j, D}^{}
\eea
One can further define the overlap functions $\Phi_{i P}$ and $\Phi_{j D}$ as~\cite{Beuthe:2001rc,Giunti:2002xg} 
\bea
\Phi_{i, P} &=& \int \frac{d^3 q }{(2\pi)^3 } \frac{d^3 k }{ (2\pi)^3 }  f_{P_i}^{} (q, Q) f_{P_f}^* (k, K) {\cal M}_{i, P}  (2\pi)^4 \delta^4 (q-k-p) \; , \\
\Phi_{j, D} &=& \int \frac{d^3 q^\prime }{(2\pi)^3 } \frac{d^3 k^\prime }{ (2\pi)^3 }  f_{D_i}^{} (q^\prime, Q^\prime) f_{D_f}^* (k^\prime, K^\prime) {\cal M}_{j, D}  (2\pi)^4 \delta^{(4)} (q^\prime-k^\prime+p)  \; ,
\eea
then transition amplitude can be rewritten as
\bea
i T_{\alpha \to \beta}^{} =  \sum_{ij} U_{\alpha  i}^* {\cal U}_{\beta j}^{} \int \frac{d^4 p}{(2\pi)^4} \Phi_i^* ( p ) \Phi_j ( p ) \Sigma^{i  j} (p) \exp \left( -i p_0 T + i \vec p \cdot L \right) \; . 
\eea
To derive the analytical formula of the neutrino oscillation probability, we  focus on two-flavor neutrino oscillation case, $\nu_e \to \nu_\mu$, where  the flavor mixing can be describe by the single angle $\theta_{}$. Integrating over the momentum $\vec p$, one has 
\bea
i T_{e \to \mu }^{} \approx  \int \frac {d^4 p }{(2\pi)^4} \Phi (p)\Phi^* (p) \left[ \frac{1}{2} \sin 2\theta \left( \Sigma^{22} -\Sigma^{11} \right) + \cos 2\theta \Sigma^{12} \right]  \exp \left( -i p_0 T + i \vec p \cdot L \right)  \nonumber \\
= \int \frac{d p^0}{ 2\pi }   \frac{(\Delta \bar m^2 /4p_0)  s_{2\theta}  + g  c_{2\theta}  } {\sqrt{\left( {\Delta \bar m^2/ 4 p_0} \right)^2 + g^2 } } \sin \sqrt{ \left( \frac{\Delta \bar m^2 }{4p_0}
\right)^2 + g^2 } L  \frac{p_0 e^{ -i  p_0 (T-L) } -p_0 e^{-i p_0 (T+L) }}{4  \pi L } |\Phi (p_0)  |^2 
\eea
The neutrino oscillation probability is the squared modulus of the amplitude  integrated over the time $T$~\cite{Beuthe:2001rc,Giunti:2002xg} 
\bea
P(\nu_e \to \nu_\mu )= \int dT \left|T_{e\to \mu}\right|^2 / (N_P N_D)
\eea
where $N_P$ and $N_D$ are normalization factors proportional $|\Phi_P|^2 $ and $|\Phi_D|^2$, respectively.  Finally one can read out the transition probability in the plane wave limit
\bea
P_{e\to \mu } =  \frac{[(\Delta \bar m^2 /4E )  s_{2\theta}  + \kappa  c_{2\theta}]^2  } {\left( {\Delta \bar m^2/ 4 E } \right)^2 + \kappa^2  } \sin^2 \left[ \sqrt{ \left( \frac{\Delta \bar m^2 }{4E}
\right)^2 + \kappa^2 } L \right]  \label{20251119-1}
\eea
which is consistent the traditional  $2 \to 2$ neutrino oscillation formula in matter.

\subsection{Comparing with the quantum mechanic result}
Neutrino oscillation probability can be derived by solving the schrodinger-like equation~\cite{Giunti:2007ry}
\bea
i \frac{d}{dx} \Psi_\alpha = \frac{1 } {2E}  \mathbb{M}^\dagger \mathbb{M} \Psi_\alpha 
\eea
where $E$ is the energy of neutrinos and $\mathbb{M}$ is effective neutrino mass matrix with $\mathbb{M} = m_0 + g \varphi $.  New mass matrix can be diagonalized by a $2\times 2$ unitary matrix ${\cal U}^\prime$ parameterized by the parameter $\theta_M$,  i.e. ${\cal U}^{\prime T} {\mathbb M} {\cal U}^\prime ={\rm Diag} \{\hat m_1, \hat m_2 \}$, with 
\bea
\tan 2\theta_M  = \frac{  \Delta m  \sin 2\theta +2 {\cal G}_{12}^{} \varphi } {  \Delta m  \cos 2\theta  + ({\cal G}_{22}- {\cal G}_{11})\varphi }
\eea
and 
\bea
\hat m_{1,2} ={1\over 2} \left[ m_{1}+ m_2 + ({\cal G}_{11}+ {\cal G}_{22})\varphi  \pm \sqrt{ [\Delta m c_{2\theta} + ({\cal G}_{22} -{\cal G}_{11})\varphi]^2 + (\Delta m s_{2\theta} + 2 {\cal G}_{12} \varphi)^2} \right]
\eea
where $\Delta $ and $\theta$ are the mass difference and the mixing angle in the vacuum. Given these parameters, the two flavor oscillation probability can be written as
\bea
P_{e\to \mu} =\sin^2 2 \theta_M \sin^2 \left( \frac{\Delta m_M^2 L }{4E} \right) \label{20251119-2}
\eea
which has the same form as the two-flavor neutrino oscillation formula in the vacuum,  up to the replacements $\theta\to \theta_M$ and $\Delta m^2 \to \Delta m^2_M$.  

Comments on the Eqs. (\ref{20251119-1})  and (\ref{20251119-2}) are listed in the following:
\begin{itemize}
\item  Eq. (\ref{20251119-1}) does not explicitly depend on the time, since we have already performed the Fourier transform to the classical field $\varphi$ when deriving propagators, while Eq. (\ref{20251119-1})  varies as time and one needs to perform time averaging on the neutrino oscillation probability in specific experimental measurement. 
\item The squared mass difference is different in two formula,  as $\tilde m_1 = m_1 + (c^2 {\cal G}_{11}+ s^2 {\cal G}_{22} -2cs {\cal G}_{12}) 2 \pi \sqrt{\rho}/m_\varphi $ and  $\tilde m_1 = m_2 + (s^2 {\cal G}_{11}+ c^2 {\cal G}_{22} + 2cs {\cal G}_{12})2 \pi \sqrt{\rho}/m_\varphi $, where $c=\cos\theta$ and $s=\sin \theta$. $\tilde m_1+ \tilde m_2 $ has the same form as $\hat m_1 + \hat m_2$.
\item Both results can be changed into the standard $2\to 2$ neutrino oscillation formula when turning off the DM-neutrino interaction.  
\end{itemize}

\begin{figure}[t]
	\centering
	\includegraphics[width=0.6\textwidth]{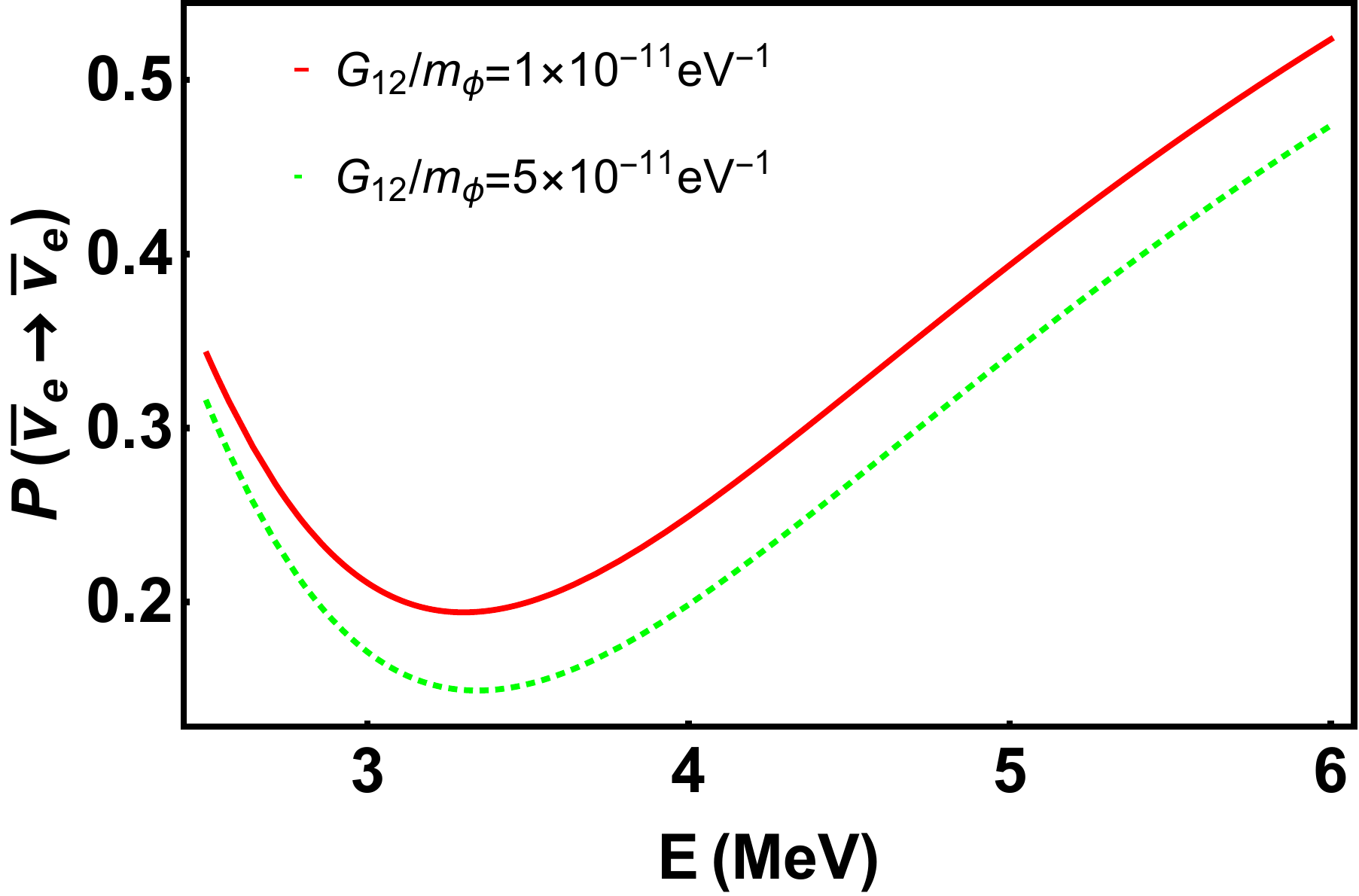}
	\caption{Neutrino oscillation probability as the function of the energy E in the JUNO.}
	\label{Juno}
\end{figure}

\subsection{Neutrino oscillations in JUNO}

Now we can perform phenomenological study to the neutrino oscillation in ultralight DM. Recently, the JUNO neutrino oscillation experiment~\cite{JUNO:2015zny} has reported their first determination of two neutrino oscillation parameters, $\sin^2 \Theta_{12} = 0.3092\pm 0.0087$ and $\Delta M_{21}^2 = (7.50 \pm 0.12) \times 10^{-5}~{\rm eV}^2$ using 59.1 days of data collected since detector completion in August 2025~\cite{JUNO:2025gmd}.  To estimate the effect induced by the DM, we set ${\cal G}_{11} ={\cal G}_{22} =0$ for simplicity, In this case, the mixing angle and the squared mass differences in the vacuum can be written as 
\bea
\theta_{12} &\approx& \Theta_{12}^{} -\frac{ 0.76 E {\cal G}_{12} \sqrt{\rho_{\rm DM}} }{m_\varphi \Delta M_{21}^2 } \\
\Delta m_{21}^2  &\approx& \Delta M_{21}^2 -\frac{32\pi^2  c_{2\theta_{12}}^2 \rho_{\rm DM} {\cal G}_{12}^2 E^2 }{m_\varphi^2 \Delta M_{21}^2 } -4\pi  s_{2\theta_{12}} (m_1 + m_2) {\cal G}_{12}^{} \frac{\sqrt{\rho_{\rm DM}^{} }}{m_\varphi}
\eea
where $E$ is the energy of neutrinos 

We further define the effective neutrino oscillation probability  as $P(E_r) =\int dE_t f( E_t, E_r, \sigma_E^{\rm exp})$ to model the experimental energy resolution, where $f$ is a Gaussian distribution function and $\sigma_E^{\rm exp}$ is the energy resolution.  We show in the Fig.~\ref{Juno}, the effective oscillation probability as the function of the energy in the JUNO experiment by assuming only $ {\cal G}_{12}$ is relevant to the oscillation.  We have set  the energy resolution $\sigma_E^{\rm exp}=0.03$ and $L=53~{\rm km}$ in  the plot,  other inputs are the same as those in the SM. 
It shows that the JUNO experiment is able to measure the DM-neutrino coupling parameters in the future.
\section{Summary and remarks}

The neutrino oscillation is an important physical phenomenon, which can not only be used to measure parameters in neutrino physics but also to explore new physics beyond the SM. In this paper, we have calculated the probability of neutrino oscillation in ultralight dark matter halos. Unlike previous approaches, we employ quantum field theory to compute this oscillation probability. We have derived the propagator of neutrinos in scalar-type ultralight DM and use it to obtain the analytical expression for the two-flavor neutrino oscillation probability. We have compared this result with that obtained through quantum mechanics. Finally, we make predictions for neutrino oscillation parameters and oscillation behavior in the JUNO experiment. It is worth mentioning that there are other types of ultralight DM, and it is necessary to study neutrino oscillation behavior in these types of DM using quantum field theory approach.

\section*{Acknowledgements}

This work was supported by the National Natural Science Foundation of China (NSFC) (Grants No. 12447105, No. 11775025 and No. 12175027), and the Fundamental Research Funds for the Central Universities.

\bibliographystyle{elsarticle-num}
\bibliography{Invariant}

\end{document}